\documentclass[11pt]{article}
\usepackage{amsmath,epsf,epsfig,float,amssymb,latexsym}
\usepackage{graphics,psfrag}
\newcommand{\vp}{{\mathbf{p}}}
\newcommand{\vq}{{\mathbf{q}}}

\newcommand{\vk}{{\mathbf{k}}}
\newcommand{\vl}{{\vec{\ell}}}

\newcommand{\Opd}{{\mathcal{O}}(p^2)}

\newcommand{\bg}{\begin{align}}
\newcommand{\eeg}{\end{align}}
\newcommand{\be}{\begin{equation}}
\newcommand{\ee}{\end{equation}}
\newcommand{\ba}{\begin{eqnarray}}
\newcommand{\ea}{\end{eqnarray}}

\newcommand{\nn}{\nonumber}

\newcommand{\barr}[1]{\not\mathrel #1}

\newcommand{\ve}{\varepsilon}

\newcommand{\vs}{\vspace{-0.2cm}}
\newcommand{\la}{\langle}
\newcommand{\ra}{\rangle}

\begin{document}

\thispagestyle{empty}

\hfill{\footnotesize HISKP-TH-09/03, FZJ-IKP-TH-2009-5}

\vspace{2cm}

\begin{center}
{\Large{\bf Chiral Effective Field Theory for Nuclear Matter with long- and
\vskip 5pt
short-range Multi-Nucleon Interactions}}
\end{center}
\vspace{.5cm}

\begin{center}
{\Large  J. A. Oller$^{a}$, A. Lacour$^{b}$  and U.-G. Mei{\ss}ner$^{b,c}$}
\vskip 10pt
{\it  $^a$Departamento de F\'{\i}sica, Universidad de Murcia, E-30071 Murcia, 
Spain}\\
{\it  $^b$Helmholtz-Institut f\"ur Strahlen- und Kernphysik (Theorie) and
  Bethe Center for Theoretical Physics\\ Universit\"at Bonn,
D-53115 Bonn, Germany}\\
{\it $^c$Institut f\"ur Kernphysik, Institute for Advanced Simulation and
J\"ulich Center for Hadron Physics\\Forschungszentrum J\"ulich, D-52425 J\"ulich,
Germany}
\end{center}

\vspace{1cm}
\noindent
\begin{abstract}
We derive a novel chiral power counting scheme for in-medium chiral
perturbation theory with explicit nucleonic and pionic degrees of freedom
coupled to external sources. It allows for a systematic expansion taking into 
account local as well as pion-mediated inter-nucleon interactions. Based on 
this power counting, one can identify classes of non-perturbative diagrams that require
a resummation. 
Within this  scheme, the pion self-energy in asymmetric nuclear matter 
is analyzed and calculated up-to-and-including next-to-leading order (NLO). It is shown
 that the corrections  involving in-medium nucleon-nucleon
interactions cancel between each other at NLO. As a result, there are
no corrections up to this order  to the linear density approximation for the
in-medium pion self-energy.
\end{abstract} 

\newpage

\section{Introduction}
\def\theequation{\arabic{section}.\arabic{equation}}
\setcounter{equation}{0}
\label{sec:int}

 One of the
long-standing  issues in nuclear physics is the calculation of atomic nuclei 
and nuclear matter properties from microscopic inter-nucleon forces in a
systematic and controlled way. This is a
non-perturbative problem involving the strong interactions.
 In the last decades, Effective Field Theory (EFT) has proven to be
an indispensable tool to accomplish such an ambitious goal. 
 The pions play a unique role in the physics of the strong interactions. They should be included consistently with the spontaneous symmetry breaking of the $SU(2)_L\otimes SU(2)_R$ chiral symmetry of the strong interactions in the limit of massless $u$ and $d$ quarks. We follow the techniques of Chiral Perturbation Theory (CHPT) \cite{wein,wein1,wein2}, with nucleons and pions as the pertinent degrees of freedom. In this approach, the pions are the Goldstone bosons associated with the spontaneous symmetry breaking of the chiral symmetry. They finally acquire a finite mass because of the small but non-vanishing masses of the lightest quarks $u$ and $d$, which explicitly break chiral symmetry. 
 For the lightest nuclear systems with two,
three and four nucleons, CHPT in nuclear systems 
 has been successfully applied 
\cite{ordo,kolck,entem,epe,epeprl,eperp,Epelbaum:2008ga}. 
 For heavier nuclei one common procedure is to employ the chiral
 nucleon-nucleon  potential derived in  CHPT combined with standard  many-body 
methods, sometimes supplied with renormalization group techniques~\cite{schaefer}.
One of the most pressing issues of
interest  is the consistent inclusion of multi--nucleon interactions involving 
three or more nucleons  in nuclear matter and nuclei, see 
e.g.~\cite{nogga,kaiser,epenm,Navratil:2007we}.

In ref.\cite{prcoller} many-body field theory was derived from quantum field
theory by considering nuclear matter as a finite density system of free nucleons at asymptotic times.
Based on these results 
ref.\cite{annp} derived a chiral power counting in the nuclear medium.\footnote{This
approach was later generalized to finite nuclei and e.g. applied to the calculation
of the pion-nucleus optical potential \cite{Girlanda:2003cq}.} However, only nucleon interactions due to pion exchanges were considered. 
In this work we derive an extended (and newly organized) power counting that takes into account local multi-nucleon interactions simultaneously to the pion-nucleon interactions. 
Many present applications of CHPT to nuclei and nuclear matter \cite{kaiser,kai1,kai2,kai3,annp,hardrock,osetdo,osetnie}, only consider
meson-baryon chiral Lagrangians (see e.g. \cite{Epelbaum:2008ga} for a summary),
without constraints from free nucleon-nucleon scattering.
In addition, as it is well known since the seminal papers of Weinberg \cite{wein1,wein2}, the nucleon propagators do
often count as the inverse of a nucleon kinetic energy $\mathcal{O}(p^{-2})$, 
so that they are much larger than assumed.
This, of course, invalidates the straightforward application to nuclear physics of 
the pion-nucleon power counting valid in vacuum,  as used e.g. 
 in refs.~\cite{annp,korean,kai1,kai2,kai3}.

All these calculations share the assumption that  spontaneous chiral symmetry breaking still holds for finite density nuclear systems. This assumption can be cross-checked  by calculating the temporal  pion decay constant in the nuclear medium \cite{annp}. Namely, let us consider the axial-vector current $A_\mu^i=\bar{q}(x)\gamma_\mu\gamma_5(\tau^i/2) q(x)$, with $q(x)$ a two-dimensional vector corresponding to the quarks fields and $\tau^i$ the Pauli matrices. Spontaneous chiral symmetry breaking results because the  axial charge, $Q_A^i=\int d^3x A_0^i(x)$, does not annihilate the ground state, denoted by  $|\Omega\ra$. 
As long as the matrix element $\la \Omega|Q_A^i|\pi^a(\vp)\ra=i (2\pi)^3 \delta(\vp) p_0\, f_t$  is not zero, the vacuum is not left invariant by the action of the axial charge and spontaneous chiral symmetry breaking happens. In the previous equation $|\pi^a(\vp)\ra$ denotes a pion state with Cartesian coordinate $a$,  three-momentum $\vp$, energy $p_0$  and $f_t$ is the temporal weak pion decay coupling.\footnote{Mathematically one can obtain meaningful results from  $i\delta(\vp)p_0$ in the chiral limit by considering wave packets \cite{annp}, $\int d\vp |\vp|^{-1} f(\vp^2)|\pi^a(\vp)\ra$ with $f(0)$=constant \cite{goldstone}.}
 Note that due to the presence of the nuclear medium one should distinguish between the spatial and temporal couplings of the pion to the axial-vector current.   The calculations in ref.\cite{annp} indicate a linear decreasing of $f_t$ with density,  $f_t=f_\pi(1-(0.26\pm 0.04)\rho/\rho_0)$, where $f_\pi=92.4~$MeV is the weak pion decay constant in vacuum and $\rho_0$ is the nuclear matter saturation density. This result clearly indicates that it makes sense to use chiral Lagrangians in the nuclear medium up to central nuclear densities. Nonetheless, a more thorough calculation of $f_t$ within our present approach, including short range nucleon-nucleon correlations, should be pursued in order to check whether the dependence in density of $f_t$ remains stable or is subject to significant corrections. On the other hand, the form of the chiral Lagrangians changes depending whether the quark condensate $\la \Omega| \bar{u}u+\bar{d}d|\Omega\ra$ is large or small. In the former case we have standard CHPT \cite{wein,bern} and in the latter the so called generalized CHPT would result \cite{jan}. In SU(2) CHPT it has been shown that the first case holds \cite{prlbern}. Refs.\cite{annp,andreas} obtain that the in-medium quark condensate decreases linearly with density  as $1-(0.35\pm 0.09)\rho/\rho_0$ \cite{annp}. Thus, the standard CHPT scenario  holds up to nuclear matter saturation density. Ref.\cite{condenk} calculated higher order corrections to this result  and found the same linear trend for symmetric nuclear matter up to $\rho \simeq \rho_0$. For higher densities,  the linear decreasing  is softened and frozen to a reduction of a 40$\%$ with respect to the vacuum value.  For the pure neutron matter  the higher order corrections calculated in \cite{condenk} do not spoil the linear decrease of the quark condensate even for densities $\rho >\rho_0$. 

Our novel power counting is applied to the problem of calculating  
the pion self-energy in asymmetric nuclear matter at next-to-leading order
(NLO).  This problem is tightly connected with that of pionic atoms
\cite{ericeric,galrep} due to the relation between the pion self-energy 
and the pion-nucleus optical potential. Despite being an old subject, 
a conclusive calculation of the pion self-energy in a systematic 
and controlled expansion is still lacking. For recent calculations see
\cite{annp,chanfray,osetdo,kwpk,korean,Girlanda:2004qa}. 
In particular, the problem of the missing S-wave repulsion, the
renormalization of the isovector scattering length
$a^-$ in the medium \cite{chanfray,osetdo} and the
energy-dependence of the isovector amplitude \cite{galrep} is not settled yet, 
despite the recent progresses \cite{weiprl,galrep,annp}.

Relativistic field theories of nuclear phenomena featuring manifest Lorentz covariance are widely used to describe properties of nuclear matter and finite nuclei. Typically nucleons are described as point Dirac-particles moving in    large isoscalar scalar and vector mean fields generated self-consistently \cite{original}. The scalar mean field drives a strong attraction of order 300-400~MeV at nuclear density and almost and equally strong vector repulsion. This is a benchmark characteristic of the so called Quantum Hadrodynamics  \cite{rev1,rev2,short1,short2}. As discussed in ref.\cite{rev2}, when the empirical low-energy nucleon-nucleon scattering is described in a Lorentz-covariant fashion, it contains strong scalar and four-vector amplitudes \cite{bdserots}. These important aspects are kept in the previous models which are widely applied for studying nuclear matter and nuclei. Applications are calculated with different degrees of refinement since the $\sigma\omega$ mean-field model of \cite{original}, including more vector and scalar fields and additional renormalizable scalar meson self-couplings \cite{rev2} with adjustable parameters. In our approach  the important nucleon-nucleon dynamics is generated by applying chiral EFT to systems with nucleons and pions, as commented above. No explicit mean fields are included, but this is not at odds with the mean-field models.  Our approach is a dynamical one that should reproduce the physical effects of such mean fields in terms of the self-interactions of the explicit degrees of freedom included.

After this introduction, we derive in section \ref{sec:pw} a novel chiral power
counting in the medium that takes into account  multi-nucleon local
interactions, pion exchanges and  the
enhancement of nucleon propagators. In
 sections \ref{sec:msby} and \ref{sec:nnself} 
we calculate the different contributions to the pion
self-energy in asymmetric nuclear matter up-to-and-including ${\cal O}(p^5)$ at NLO. 
Interestingly, we show in section \ref{sec:nnself} that the
different NLO contributions from nucleon-nucleon scattering cancel between each
other.  Conclusions are given in section \ref{sec:conc}.

\section{Chiral Power Counting}
\def\theequation{\arabic{section}.\arabic{equation}}
\setcounter{equation}{0}
\label{sec:pw}

In ref.\cite{prcoller} the effective chiral pion Lagrangian was determined in
the nuclear medium in the presence of external sources. For that the Fermi
seas of protons and neutrons were integrated out making use of functional
techniques. A similar approach was followed in ref.\cite{sainio} but for the
case of only one nucleon. In this way it is manifestly shown that
pion or nucleon field redefinitions do not affect physical
observables also in nuclear matter because they appear as integration variables 
 in a functional. Nonetheless, in ref.\cite{prcoller} only the meson-baryon
chiral Lagrangian is employed. More precisely, if we write a general chiral Lagrangian
in terms of an increasing number of nucleon fields $\psi$,
\begin{align}
{\cal L}_{\rm eff}={\cal L}_{\pi\pi}+{\cal L}_{\bar{\psi}\psi}
+{\cal L}_{\bar{\psi}\bar{\psi}\psi\psi}+\ldots
\label{lagnn}
\end{align}
 only the contributions from  ${\cal L}_{\pi\pi}$ and ${\cal
L}_{\bar{\psi}\psi}$ were retained in  ref.\cite{prcoller}. Based on these results, the authors of
ref.\cite{annp}  derived a chiral power counting in  the nuclear medium.

\begin{figure}[t]
\psfrag{q1}{$q$}
\psfrag{l}{$i$}
\psfrag{m}{$j$}
\centerline{\epsfig{file=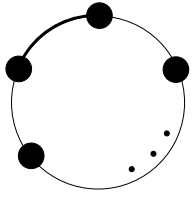,width=0.14 \textwidth,angle=0}}
\vspace{0.2cm}
\caption[pilf]{\protect \small
Representation of an IGV. See the text for further details.
\label{fig:mgv}}
\end{figure}

\begin{figure}[t]
\centerline{\epsfig{file=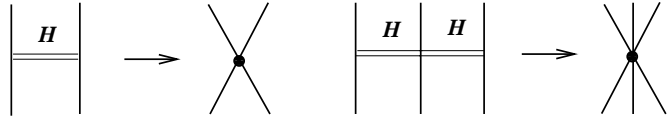,width=0.50 \textwidth,angle=0}}
\vspace{0.2cm}
\caption[H]{\protect \small
Representation of multi-nucleon interactions through the multiple exchange
of heavy mesons $H$ as described in the text.
\label{fig:Hfields}}
\end{figure}

Ref.\cite{prcoller}  establishes the concept of an ``in-medium generalized vertex'' (IGV).  
Such type of vertices result because
one can connect several bilinear vacuum vertices through the exchange of baryon
propagators with the flow through the loop of one unit of baryon number,
contributed by the nucleon Fermi seas. This is schematically shown in
fig.\ref{fig:mgv} where the thick arc segment indicates an insertion of a
Fermi sea. At least one is needed because otherwise we would have a vacuum closed
nucleon loop that in a low energy effective field theory  is buried in the 
 higher order chiral counterterms. On the other hand, a filled large circle in
fig.\ref{fig:mgv} indicates a bilinear nucleon vertex from ${\cal L}_{\pi N}$,
while the dots refer to the insertion of any number of them.\footnote{
${\cal L}_{\pi N}^{(i)}$ corresponds to the CHPT pion-nucleon Lagrangian of chiral order $i$.} 
 It was also stressed in ref.\cite{annp} that within a nuclear environment a nucleon
 propagator could have a ``standard'' or ``non-standard'' chiral counting. To see
 this note that a soft momentum $Q \sim p$, related to pions or external sources
 attached to the bilinear vertices in fig.\ref{fig:mgv}, can be associated to any of the
 vertices.
 Denoting by $k$ the on-shell four-momenta associated with one
Fermi sea insertion in the IGV, the four-momentum
running through the $j^{th}$ nucleon propagator can be written as $p_j=k+Q_j$.
In this way, 
\begin{align}
i\frac{\barr{k}+\barr{Q}_j+m}{(k+Q_j)^2-m^2+i\epsilon}=
 i \frac{\barr{k}+\barr{Q}_j+m}{Q_j^2+2 Q_j^0 E(\vk) -2 {\mathbf{Q}}_j
 \mathbf{k}+i\epsilon}~,
\label{pro.1}
\end{align}
 where  $E(\vk)=\vk^2/2m$, with $m$ the physical nucleon mass (not the bare
one), and $Q_j^0$ is the temporal component of $Q_j$. We have just shown in the previous equation the free part of an in-medium nucleon propagator because this is enough for our present discussion.  
Two different situations occur depending on the value of $Q_j^0$. If $Q_j^0={\cal O}(m_\pi)={\cal O}(p)$  one has the standard counting so that the chiral expansion of the propagator in eq.(\ref{pro.1}) is
\begin{align}
i \frac{\barr{k}+\barr{Q}_j+m}{2 Q^0_j m+i\epsilon} 
 \left(1-\frac{Q_j^2-2 {\mathbf{Q}}_j \cdot \mathbf{k}}{2Q_j^0 m}+\Opd \right)~.
\end{align}
Thus, the baryon propagator  counts as a quantity of ${\cal O}(p^{-1})$. But it could also
occur that $Q_j^0$ is  of the order of a kinetic nucleon energy in the nuclear
medium or that it even vanishes. The dominant term in eq.(\ref{pro.1}) is then
\begin{align}
-i\frac{\barr{k}+\barr{Q}_j+m}{\mathbf{Q}^2_j+2\mathbf{Q}_j\cdot\vk-i
\ve} ~,
\end{align}
and the nucleon propagator should be counted as ${\cal O}(p^{-2})$, instead
of the previous ${\cal O}(p^{-1})$. This is referred to as the ``non-standard'' 
case in ref.\cite{annp}. We should  stress that this situation also occurs 
already in the vacuum when considering the two-nucleon reducible diagrams in 
nucleon-nucleon scattering. This is indeed the reason advocated in
ref.\cite{wein1} 
for solving a Lippmann-Schwinger equation with the nucleon-nucleon potential 
given by the two-nucleon irreducible diagrams. 
In the present investigation, we extend the results of refs.\cite{prcoller,annp} in a twofold way.
i) We are able to consider chiral Lagrangians
with an arbitrary number of baryon fields (bilinear, quartic, etc). 
First only
bilinear vertices like in refs.\cite{prcoller,annp} are considered,  but now  the additional 
exchanges of  heavy meson fields of
any type are allowed. The latter  should be considered as
merely auxiliary fields that allow one to find a tractable
representation of the multi-nucleon interactions that result when  the
masses of the heavy mesons tend to infinity.\footnote{Such methods are
also used in the so-called nuclear lattice simulations, see e.g. \cite{Borasoy:2006qn}.} 
 These heavy meson fields are 
denoted in the following by $H$, see fig.\ref{fig:Hfields}, 
and a heavy meson propagator is counted as ${\cal O}(p^0)$ due to their large masses.
 ii) We take the non-standard counting from the start
and  
 any nucleon propagator is considered 
 as ${\cal O}(p^{-2})$.
In this way, no diagram whose chiral order is actually lower
than expected if the nucleon propagators were counted assuming the standard rules is lost. This is a
novelty in the literature.

In the following $m_\pi\sim k_F\sim {\cal O}(p)$ are taken of the same chiral order,  and are considered  much smaller than a hadronic scale $\Lambda_\chi$  of several hundreds of MeV that results by integrating out all other particle types, including nucleons with larger three-momentum, heavy mesons and nucleon isobars \cite{wein2}. 
The chiral order of a given diagram is represented by $\nu$ and is given by
\be
\nu=4L_H+4L_\pi-2I_\pi+\sum_{i=1}^V d_{i}-\sum_{i=1}^{V_\rho}2m_i
+\sum_{i=1}^{V_\pi}\ell_i
+\sum_{i=1}^{V_\rho}3~.
\label{count}
\ee
From left to right,  $L_H$ is the number of loops due to  the internal heavy mesonic lines, $L_\pi$ that  of
  pionic loops and $I_\pi$ is the number of internal pionic lines. Each loop introduces a factor of four in the power counting, because of the integration over the free four-momentum, and a pion propagator reduces the order by two units. The quantity $d_i$ is the chiral order of the $i^{th}$ bilinear vertex in the baryonic fields and $V$ is the total number of such vertices.   $V_\rho$ is the number of IGVs  and $m_i$ is the number of 
 nucleon propagators in $i^{th}$  IGV minus one, and every of them reduces by two units the chiral counting as discussed previously. The definition of $m_i$ contains the removal of one baryon propagator because there is always at least one  Fermi sea insertion for each IGV that increases the chiral counting in three units because the associated integration over a Fermi sea, $\int d^3 k \theta(\xi_i-|\vk|)$, with $\xi_i$ the corresponding Fermi momentum. This fact gives rise to the last sum in eq.\eqref{count}.  Other symbols that appear are
$\ell_i$, that is the chiral order of a vertex without baryons (only
  pions and external sources), and $V_\pi$ which is  the total number of the latter ones. We have not included in eq.(\ref{count}) any contribution from $\pi$-$H$
vertices without baryons because in the limit when the mass of the $H$
fields is taken to infinity the $H$ propagators are contracted to a point
and the pions will be  always attached to baryons.

 Let us note that associated with the bilinear vertices in an IGV one has  four-momentum conservation Dirac delta functions that can be used to fix the momentum of each of the
 baryonic lines joining them, except one for the running three-momentum due to the Fermi sea insertion. 
Let us now introduce another symbol, $V_\Phi$. Here, we take as a whole any set of IGVs  that are joined through {\it heavy} mesonic lines $H$, whose total number is $I_H$. The number of these clusters of in-medium 
generalized vertices is denoted by
$V_\Phi$.  In this way, we can write 
\be
L_H=I_H-\sum_{i=1}^{V_\Phi}\left(V_{\rho,i}-1\right)=I_H-V_\rho+V_\Phi~,
\label{lh}
\ee
where $V_{\rho,i}$ is the number of IGVs within the $i^{th}$ set of generalized vertices
connected by heavy mesonic lines. Additionally, they could be connected between them or with other IGVs belonging to other clusters by pionic lines. Since there is a total four-momentum conservation delta function associated
to every of these clusters it follows that 
\be
L_\pi=I_\pi-V_\pi-V_\Phi+1~.
\label{lpi}
\ee
 
\begin{figure}[t]
\psfrag{CLUSTER 1}{Cluster 1}
\psfrag{CLUSTER 2}{Cluster 2}
\centerline{\epsfig{file=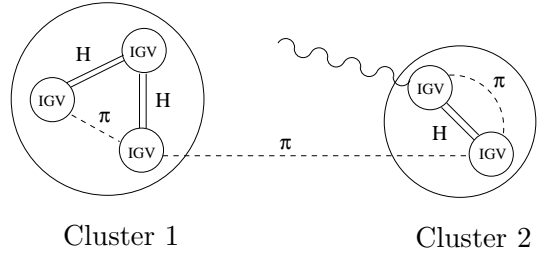,width=0.40 \textwidth,angle=0}}
\vspace{0.2cm}
\caption[H]{\protect \small
Representation of a possible arrangement of IGVs separated in two clusters. In this figure $V_\rho=5$, $V_\Phi=2$, $I_\pi=3$, $I_H=3$ and $E=1$. The pions are indicated by the dashed lines and the external source by a wavy line. Eqs.(\ref{lpi}) and (\ref{lh}) imply that $L_\pi=2$ and $L_H=0$ as it should.
\label{fig:aspect}}
\end{figure}

These relations are illustrated in fig.\ref{fig:aspect} where a possible arrangement  of IGVs is shown. On the other hand,
\be
2I_H+2I_\pi+E=\sum_{i=1}^V v_i+\sum_{i=1}^{V_\pi}n_i~.
\label{lines}
\ee
Here,  $v_i$ is the number of mesonic lines attached to the 
$i^{th}$ bilinear vertex, $n_i$ is the number of pions in the $i^{th}$ mesonic vertex and $E$ is the 
number of external pions. 
Taking into account eqs.(\ref{lh}), (\ref{lpi}) and eq.(\ref{lines}),  eq.(\ref{count}) reads~,
\be
\nu=2I_H-E+4-4V_\pi+\sum_{i=1}^{V_\pi}(\ell_i+n_i)+\sum_{i=1}^V (d_i+v_i)-2 m -V_\rho~,
\label{int}
\ee
with $m=\sum_{i=1}^{V_\rho}m_i~.$ 
We now employ in eq.(\ref{int}) that
 $V_\rho+ m=V~,$ and $2 I_H=\sum_{i=1}^V \omega_i$~,
where  $\omega_i$ is the number of heavy meson internal lines for the $i^{th}$ bilinear vertex. Then, 
we arrive at our final equation
\be
\nu=4-E+\sum_{i=1}^{V_\pi}(n_i+\ell_i-4)+\sum_{i=1}^V(d_i+\omega_i-1)+
\sum_{i=1}^m(v_i-1)+\sum_{i=1}^{V_\rho} v_i~.
\label{fff}
\ee
Note that $\nu$ given in eq.(\ref{fff}) is bounded from below because
$
n_i+\ell_i-4\geq 0$, as $\ell_i\geq 2$ and $n_i\geq 2$, except for a finite number of terms that could contain only one pion line but always having external sources attached to them. Similarly 
$d_i+\omega_i-1\geq 0$.
For the  pion-nucleon Lagrangians this is always true as $d_i\geq 1$. For those
bilinear vertices mediated
by heavy lines  $d_i\geq 0$ but then $\omega_i\geq 1$.  For the term before the last one in eq.(\ref{fff}) $v_i-1\geq 0$,
except for the higher-order nucleon-mass renormalization counter terms or  the finite number of terms which would not have pionic lines but
only external sources  from ${\cal L}_{\pi N}$. The former terms have
$d_i\geq 2$ and then  $(d_i+\omega_i-1)+(v_i-1)\geq 0$. For $d_i=2$
the chiral order does not increase but these terms can be
absorbed in the physical nucleon mass. For the
last term in eq.(\ref{fff}) $v_i\geq 0$ and then it 
is positive.  It is worth stressing 
 that adding
 a new IGV to a connected diagram increases the counting at least by 
 one unit because then $v_i\geq 1$. Using again that $V_\rho+m=V$ eq.\eqref{fff} can be rewritten as
 \be
\nu=4-E+\sum_{i=1}^{V_\pi}(n_i+\ell_i-4)+\sum_{i=1}^V(d_i+\omega_i+v_i-2)+V_\rho~.
\label{fff2}
\ee

The number $ \nu$ given in eq.(\ref{fff}) represents a lower bound for the actual chiral power of a
diagram, $\mu$, so that $\mu\geq \nu$. The real chiral order of a diagram might be different from $\nu$  because 
the nucleon propagators are counted always as ${\cal O}(p^{-2})$ to obtain eq.(\ref{fff}), while for some diagrams there could be propagators  that follow the standard counting. 
 Eq.(\ref{fff}) implies the following conditions for
augmenting the number of lines in a diagram without increasing the chiral power
by:
\begin{enumerate}
\item adding pionic lines attached to mesonic vertices, $\ell_i=n_i=2$.
\item adding pionic lines attached to meson-baryon vertices, $d_i=v_i=1$.
\item adding heavy mesonic lines attached to bilinear vertices, $d_i=0$, $\omega_i=1$.
\end{enumerate}

There is no way to decrease the order.\footnote{Only by  adding vertices with
$\ell_i=2$ and $n_i<2$ or $d_i=1$  and $v_i=0$. However, its 
 number  is bounded from above by the necessarily finite number of  external
sources.}
We apply eq.(\ref{fff})  by increasing step by step $V_\rho$ up to the order
considered. For each $V_\rho$ we look for those 
 diagrams that do not increase the order according to  the previous
list. Some of these diagrams are indeed of higher order and one can refrain from
calculating them by  establishing which of the nucleon propagators 
 scale as ${\cal O}(p^{-1})$. 
 
\begin{figure}[t]
\psfrag{Vr=1}{{\small $V_\rho=1$}}
\psfrag{Vr=2}{{\small $V_\rho=2$}}
\psfrag{Op4}{{\small ${\cal O}(p^4)$}}
\psfrag{Op5}{{\small $ {\cal O}(p^5)$}}
\psfrag{i}{{\small $i$}}
\psfrag{j}{{\small $j$}}
\psfrag{q}{{\small $q$}}
\centerline{\fbox{\epsfig{file=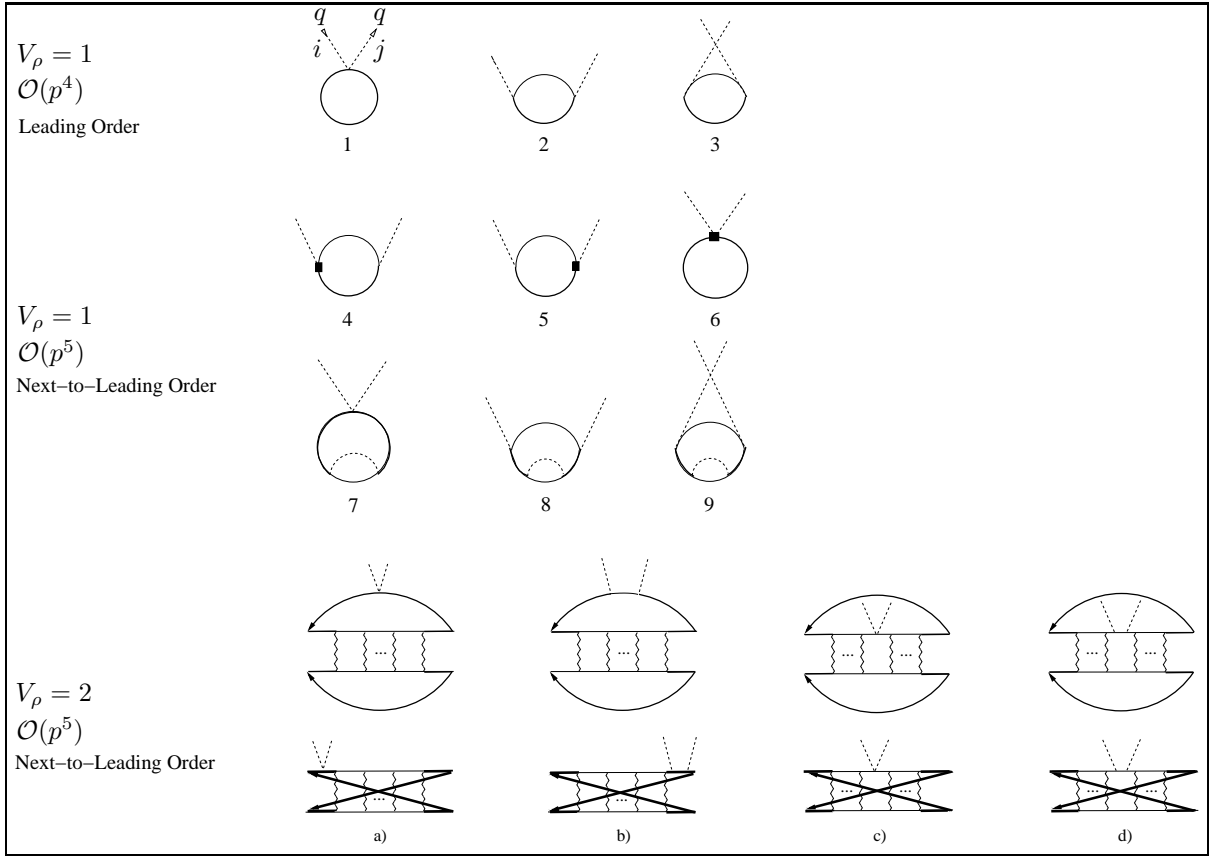,width=.9\textwidth,angle=0}}}
\vspace{0.2cm}
\caption[pilf]{\protect \small
Contributions to the in-medium pion self-energy  up
to  NLO or ${\cal O}(p^5)$. The pions are indicated by the dashed lines and
the squares correspond to NLO pion-nucleon vertices. The wiggly lines correspond to  the
nucleon-nucleon interaction kernel, see fig.\ref{fig:nnself}a, 
that  is iterated as meant by the ellipsis.
\label{fig:all}}
\end{figure}

\section{Meson-baryon contributions to the pion self-energy}
\label{sec:msby}
\def\theequation{\arabic{section}.\arabic{equation}}
\setcounter{equation}{0}

Let us apply  the chiral counting given in eq.(\ref{fff2}) 
 to
calculate the pion self-energy in the nuclear medium up to NLO or ${\cal
O}(p^5)$, with the different contributions
shown in fig.\ref{fig:all}. 
The nucleon propagator, $G_0(k)_{i_3}$, contains both the free and the in-medium piece \cite{fetter}, \be
G_0(k)_{i_3}=
\frac{\theta(\xi_{i_3}-|\vk|)}{k^0-E(\vk)-i\epsilon}+\frac{\theta(|\vk|-\xi_{i_3})}{k^0-E(\vk)+i\epsilon}=
\frac{1}{k^0-E(\vk)+i\epsilon}+i(2\pi)\theta(\xi_{i_3}-|\vk|)\delta(k^0-E(\vk))~.
\label{nuc.pro}
\ee
In this equation the subscript $i_3$ refers to the type of nucleon, 
 so that, $i_3=p$~ corresponds to the proton and 
$i_3=n$ to the neutron. 
 Our convention for the pion self-energy,  $\Sigma$, is such that the dressed
pion propagator reads
\begin{align}
 \Delta_\pi (q) = \frac{1}{q^2-m_\pi^2+\Sigma}~.
\end{align} 

We employ the pion-nucleon Heavy Baryon CHPT (HBCHPT)  Lagrangians at ${\cal O}(p)$ and ${\cal O}(p^2)$, that can be found e.g. in refs.\cite{nnk,ulfrev}. For completeness we reproduce them,
\begin{align}
{\cal L}_{\pi N}^{(1)}&=\bar{N}(i  D_0-\frac{g_A}{2}\vec{\sigma}\cdot \vec{u})N~,\nn\\
{\cal L}_{\pi N}^{(2)}&=\bar{N}\Biggl(
\frac{1}{2M}\vec{D}\cdot \vec{D}+i\frac{g_A}{4M}\left\{ \vec{\sigma}\cdot \vec{D},u_0\right\}+2c_1 m_\pi^2(U+U^\dagger)+\left(c_2-\frac{g_A^2}{8M}\right)u_0^2+c_3 u_\mu u^\mu
\Biggr)+\ldots
\label{lags}
\end{align}
where the ellipses represent terms that are not needed here.
In this equation, $N$ is the two component field of the nucleons,  $g_A$ is the axial pion-nucleon coupling and $D_\mu=\partial_\mu+\Gamma_\mu$ the covariant chiral derivative, being $\Gamma_\mu=[u^\dagger, \partial_\mu u]$.     The pion fields $\vec{\pi}(x)$ enter in the matrix $u=\exp(i\vec{\tau}\cdot \vec{\pi}/2f)$, in terms of which $u_\mu=i\left\{u^\dagger,\partial_\mu u\right\}$ and $U= u^2$,  with $f$ the weak pion decay constant in the SU(2) chiral limit. The $c_i$ are chiral low energy constants whose values are fitted from phenomenology \cite{ulfrev}.

The leading contribution to the pion self-energy corresponds to the diagrams 1--3 on fig.\ref{fig:all}. 
The diagram 1 results by closing the Weinberg-Tomozawa pion-nucleon interaction
(WT), obtained by expanding ${\cal L}_{\pi N}^{(1)}$ in eq.\eqref{lags} up to two pion fields, 
\begin{align}
\Sigma_1&=\frac{-i q^0}{2f^2}\varepsilon_{i j 3}(\rho_p-\rho_n)~,
\label{eq:sig1}
\end{align}
where  the proton(neutron) density is given by
 $
\rho_{p(n)}=\xi_{p(n)}^3/3\pi^2.$
$\Sigma_1$  is then a S-wave isovector self-energy.  
The sum of the diagrams 2 and 3 of fig.\ref{fig:all}  is
\begin{align}
\Sigma_2&=\frac{ig_A^2\,\vq^2}{2f^2 q^0}\varepsilon_{i j 3}(\rho_p-\rho_n)-\frac{g_A^2}{4f^2}\frac{(\vq^2)^2}{m q_0^2}\delta_{i j}(\rho_p+\rho_n)~.
\label{eq:sig2}
\end{align}
They involve the one-pion vertex from ${\cal L}_{\pi N}^{(1)}$, which is proportional to $g_A$. 
This is a P-wave self-energy where  the first term  is isovector and the       second  is isoscalar. 

 Now, we move to the NLO contributions. The vertices from ${\cal L}_{\pi N}^{(2)}$ in eq.\eqref{lags} are indicated by squares in the fig.\ref{fig:all}. 
 It should be understood that the pion lines can leave or
enter the diagrams  4 and 5 of fig.\ref{fig:all}. 
The sum of these two diagrams gives the result,
\be
\Sigma_3=\frac{g_A^2 \vq^2}{2m f^2}(\rho_p+\rho_n)\delta_{ij}~.
\label{eq:sig3}
\ee
This is a P-wave isoscalar contribution that stems from the term linear in $g_A$ in ${\cal L}_{\pi N}^{(2)}$, that is a recoil correction to  ${\cal L}_{\pi N}^{(1)}$. 
 The diagram 6 of fig.\ref{fig:all}  is given by
\be
\Sigma_4=\frac{-2\delta_{i j}}{f^2}\left(
2 c_1\,m_\pi^2-q_0^2 (c_2+c_3-\frac{g_A^2}{8 m})+c_3\,\vq^2\right)(\rho_p+\rho_n)~,
\label{sigma4}
\ee
in terms of the $c_i$ couplings of  ${\cal L}_{\pi N}^{(2)}$, eq.\eqref{lags}.
 This is an isoscalar contribution where  the term 
$-2\delta_{ij}c_3\vq^2(\rho_p+\rho_n)/f^2$ is P-wave  and the rest is S-wave.

Next, let us consider the contributions to the pion self-energy due to the nucleon
self-energy from a one-pion loop as depicted in the diagrams 7--9 of fig.\ref{fig:all}, with vertices calculated from ${\cal L}_{\pi N}^{(1)}$. The diagrams originate by dressing  the in-medium nucleon
propagator of the diagrams 1--3  by the one-pion loop
nucleon self-energy, 
\be
\Sigma^\pi=\frac{1+\tau_3}{2}\Sigma_p^\pi+\frac{1-\tau_3}{2}\Sigma_n^\pi~,
\label{sig.pi.def}
\ee
with $\Sigma_p^\pi$ and $\Sigma_n^\pi$ the proton and nucleon self-energies
due to the in-medium pion-nucleon loop. Their values at  $k^0=0$  are subtracted because we employ the physical nucleon mass.
  The contribution from the diagram 7 of fig.\ref{fig:all} can be written as
\begin{align}
\Sigma_5&=\frac{q^0}{f^2}\varepsilon_{i j 3}
\int\frac{d^4k}{(2\pi)^4}\left(G_0(k)^2_p \Sigma_p^\pi-G_0(k)^2_n \Sigma_n^\pi\right)e^{ik^0\eta}~,
\label{sig5.1}
\end{align}
where $e^{ik^0\eta}$, $\eta\to 0^+$, is the convergence factor associated to any closed loop made up by a single nucleon line \cite{fetter}. 
 Taking into account that 
 \begin{align}
 G_0(k)^2_{i_3}=-\partial G_0(k)_{i_3}/\partial k^0~,
\label{pro.sq} \end{align}
  we  then integrate by parts, which is possible thanks to the convergence factor. It results,
 \begin{align}
 \Sigma_5=\frac{q^0}{f^2}\varepsilon_{i j 3}\int\frac{d^4 k}{(2\pi)^4}\left(
 G_0(k)_p\frac{\partial \Sigma_p^\pi}{\partial k^0}-G_0(k)_n \frac{\partial         \Sigma_n^\pi}{\partial k^0}
 \right)e^{ik^0\eta}~.
\label{fin.sig.5}
 \end{align}
 Since $\eta\to 0$ the additional term obtained by taking the derivative of  $e^{i k^0\eta}$ with respect to $k^0$ in eq.\eqref{sig5.1} does not contribute and is not shown. 
 $\Sigma_5$  is  an isovector S-wave pion self-energy contribution. For the diagrams 8 and 9 of the same figure one has analogously
\begin{align}
\Sigma_6&=-\frac{g_A^2}{f^2}\frac{\vq^2}{q^0}\varepsilon_{i j 3}
\int\frac{d^4k}{(2\pi)^4}\left(
G_0(k)_p \frac{\partial \Sigma^\pi_{p}}{\partial k^0} -
G_0(k)_n \frac{\partial \Sigma^\pi_{n}}{\partial k^0}
\right)e^{ik^0\eta}\nn\\
&+\frac{ig_A^2}{f^2}\frac{\vq^2}{q_0^2}\delta_{i j}
\int\frac{d^4k}{(2\pi)^4}\left(\
 G_0(k)_p \Sigma^\pi_{p} 
+
 G_0(k)_n \Sigma^\pi_{n}
\right)e^{ik^0\eta}~.
\label{sig.6.f1}
\end{align}
$\Sigma_6$ is a P-wave self-energy contribution but while the first line is of
isovector character, the one in the second line is
isoscalar. This last term is indeed a NNLO or
${\cal O}(p^6)$ contribution because the pion-loop nucleon self-energy is
${\cal O}(p^3)$  and we neglect it.
 The free pion-loop nucleon self-energy is calculated in HBCHPT~\cite{ulfrev}, employing dimensional regularization in the $\overline{MS}-1$ scheme. Its derivative is
\begin{align}
\frac{\partial\Sigma_{p(n),f}^\pi}{\partial k^0}=\frac{3 g_A^2}{32\pi^2
f^2}\left[
m_\pi^2+k_0^2-3 k^0 \sqrt{b}\left(
i\log\frac{k^0+i\sqrt{b}}{-k^0+i\sqrt{b}}+\pi
\right)
\right]~,
\label{d.sif}
\end{align}
with $b=m_\pi^2-k_0^2-i\eta~.$
Hence, because $\partial\Sigma_{p(n),f}^\pi/\partial k^0={\cal O}(p^2)$ when
inserted in $\Sigma_5$ and $\Sigma_6$ it gives rise to an ${\cal O}(p^6)$
contribution that we neglect in the present work.
 The in-medium contribution to the pion-loop nucleon self-energy involves the finite 
integral 
\begin{align}
I_m=2\pi \int\frac{d^4\ell}{(2\pi)^4}\frac{{\vl}^2 \delta(k^0-\ell^0)
\theta(\xi_{i_3}-|\vk-\vl|)}{\ell^2-m_\pi^2+i\epsilon}=-\int
\frac{d^3\ell}{(2\pi)^3}
\frac{{\vl}^2 \theta(\xi_{i_3}-|\vk-\vl|)}{b+\vl^2-i\epsilon}~.
\label{im1.def}
\end{align}
  This integral only  depends on $k^0$  through the variable $b=m_\pi^2-k_0^2$.
Since $\partial I_m/\partial k^0=-2k^0\partial I_m/\partial b={\cal O}(p^2)$, 
because $k^0={\cal O}(p)$, the in-medium part of the pion loop
contribution to the nucleon self-energy  gives rise to ${\cal O}(p^6)$ terms for the pion self-energy. As a
result, $\Sigma_5$ and $\Sigma_6$ are at least ${\cal O}(p^6)$.
 Notice that 
from eq.(\ref{fff}) these contributions were firstly booked as ${\cal O}(p^5)$ because it was considered that 
$\partial \Sigma^\pi/\partial k^0={\cal O}(p)$ as $\Sigma^\pi={\cal O
}(p^3)$ and $k^0={\cal O}(p^2)$. But since for these diagrams  $V_\rho=1$, with only one closing nucleon line, $k^0={\cal O}(p)$.


\section{In-medium nucleon-nucleon scattering contributions}
\label{sec:nnself}
\setcounter{equation}{0}
We now consider those NLO contributions to the pion self-energy in the nuclear medium that involve the nucleon-nucleon interactions. 
They are depicted in the diagrams of the last two rows of fig.\ref{fig:all}, where the ellipsis indicate the iteration of the two-nucleon reducible loops. 
For the diagrams b) and d) of fig.\ref{fig:all}
the pion lines can leave or enter the diagrams. 
It is remarkable that these NLO contributions cancel between each
other. 
On the other hand, since
$V_\rho=2$ in these contributions one needs only the nucleon-nucleon scattering
amplitude at ${\cal O}(p^0)$ to match with our required precision at NLO. 
This amplitude is obtained by iterating in  an infinite ladder of two nucleon reducible loops,  with full in-medium nucleon propagators,  the tree level amplitudes obtained from the ${\cal O}(p^0)$ Lagrangian with four nucleons \cite{wein2} and from the one-pion exchange with the lowest order
pion-nucleon coupling. This  procedure would correspond in vacuum to the leading nucleon-nucleon scattering amplitude according to refs.\cite{wein1,wein2}. 
 
\begin{figure}[t]
\psfrag{k}{$k$}
\psfrag{p}{$p$}
\psfrag{l}{$\ell$}
\psfrag{pi}{$\pi$}
\psfrag{r}{$k-\ell$}
\centerline{\epsfig{file=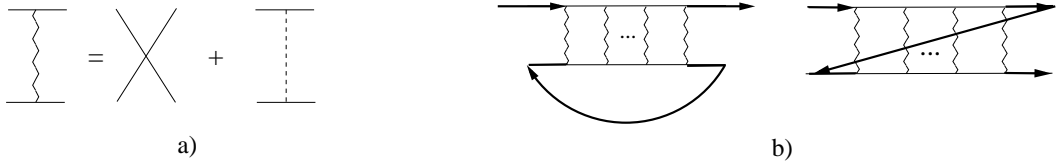,width=.8\textwidth,angle=0}}
\vspace{0.2cm}
\caption[pilf]{\protect \small
a) The wiggly line represents the sum of the
 leading one-pion exchange and  four-nucleon contact interactions, $T_{NN}^c+T_{NN}^{1\pi}$, eqs.\eqref{feynman} and \eqref{1pi.gen}. b) In-medium nucleon self-energy due to the nucleon-nucleon interactions with the Fermi seas.
\label{fig:nnself}}
\end{figure} 

The ${\cal O}(p^0)$ lowest order four nucleon Lagrangian \cite{wein2} is 
\be
{\cal L}_{NN}^{(0)}=-\frac{1}{2}C_S (\overline{N}N)(\overline{N}N)
-\frac{1}{2}C_T(\overline{N}\vec{\sigma} N)(\overline{N}\vec{\sigma} N)~.
\label{lnn}
\ee
Of course, this Lagrangian only contributes to the  S-wave nucleon-nucleon scattering. The scattering amplitude for the  process $N_{s_1,i_1}(\vp_1) N_{s_2,i_2}(\vp_2) \to  N_{s_3,i_3}(\vp_3) N_{s_4,i_4}(\vp_4)$, with $s_m$ a spin label and $i_m$ an isospin one, that follows from the previous Lagrangian is
\begin{align}
T_{NN}^{c}&=-C_S\left(\delta_{s_3 s_1}\delta_{s_4 s_2}\,\delta_{i_3i_1} \delta_{i_4 i_2}-\delta_{s_3 s_2}\delta_{s_4 s_1}\,\delta_{i_3i_2} \delta_{i_4i_1}\right)\nn\\
&-C_T\left(\vec{\sigma}_{s_3 s_1}\cdot \vec{\sigma}_{s_4 s_2}\, \delta_{i_3i_1}\delta_{i_4i_2}-\vec{\sigma}_{s_3 s_2}\cdot \vec{\sigma}_{s_4 s_1} \,\delta_{i_3i_2}\delta_{i_4i_1}\right)~.
\label{feynman}
\end{align}

Regarding the one-pion exchange at leading order, with vertices from   ${\cal L}_{\pi N}^{(1)}$, one has the expression
\begin{equation}
T_{NN}^{1\pi} = \frac{g_A^2}{4f^2}\left[
\frac{
(\vec{\tau}_{i_3i_1}\cdot \vec{\tau}_{i_4i_2})
(\vec{\sigma}\cdot \vq)_{s_3s_1}(\vec{\sigma}\cdot \vq)_{s_4s_2}}{\vq^2+m_\pi^2-i\epsilon}
\right.
- \left.\frac{(\vec{\tau}_{i_4i_1}\cdot 
\vec{\tau}_{i_3i_2})(\vec{\sigma}\cdot \vq')_{s_4s_1}(\vec{\sigma}\cdot \vq')_{s_3s_2}}{{\vq'}^2+m_\pi^2-i\epsilon}
\right]~,
\label{1pi.gen}
\end{equation}
with $\vq=\vp_3-\vp_1$ and $\vq'=\vp_4-\vp_1$.
 In the following, the sum $T_{NN}^{c}+T_{NN}^{1\pi}$ is represented diagrammatically  by the
exchange of a wiggly line,  as in fig.\ref{fig:nnself}a.

The diagrams a) and c) of fig.\ref{fig:all} involve the WT
vertex while b) and d) contain the pole terms of pion-nucleon
scattering. At leading order in the chiral counting the sum of the latter two has the same structure as the WT term, with the resulting vertex given by
\begin{align}
-\frac{iq^0}{2f^2}\left(1-g_A^2\frac{\vq^2}{q_0^2}\right) \varepsilon_{ijk}\tau^k~.
\label{ver:wte}
\end{align}
We can then discuss simultaneously all the diagrams in the last two rows  of fig.\ref{fig:all} employing the latter vertex.  The sum of the diagrams a) and b) of fig.\ref{fig:all}  can be written in terms of the
nucleon self-energy in the nuclear medium due to the nucleon-nucleon scattering. It reads
\begin{align}
\Sigma_7&=\frac{q^0}{2f^2}\left(1-g_A^2\frac{\vq^2}{q_0^2}\right)
\varepsilon_{ijk}\int\frac{d^4k_1}{(2\pi)^4} e^{ik_1^0\eta}
\hbox{ Tr}\left\{
\tau^k\left(\frac{1+\tau_3}{2}G_0(k_1)_p+\frac{1-\tau_3}{2}G_0(k_1)_n
\right)\Sigma_{NN}\left(\frac{1+\tau_3}{2}G_0(k_2)_p\right.\right.\nn\\
&\left.\left.+\frac{1-\tau_3}{2}G_0(k_2)_n
\right) \right\}~.
\label{sel.nn}
\end{align}
Here,
\be
\Sigma_{NN}=\frac{1+\tau_3}{2}\Sigma_{p,NN}+\frac{1-\tau_3}{2}\Sigma_{n,NN}~,
\ee
with $\Sigma_{p(n),NN}$ the proton (neutron) self-energy in the nuclear medium due to the nucleon-nucleon interactions, fig.\ref{fig:nnself}b,
 \be
\Sigma_{i_3,NN}=-i\sum_{\alpha_2,\sigma_2}\int\frac{d^4 k_2}{(2\pi)^4}G_0(k_2)_{\alpha_2} T_{NN}(k_1 \sigma_1 i_3,  k_2  \sigma_2 \alpha_2|k_1 \sigma_1 i_3,  k_2  \sigma_2 \alpha_2) e^{ik_2^0\eta}~.
 \label{sel.n}
 \ee
 In this expression $T_{NN}$ is the nucleon-nucleon scattering amplitude between the indicated
initial and final off-shell states. These are characterized by three labels. The first label corresponds to the four-momentum, the second to the spin and the third 
to the isospin. Note that in the equation there is a sum over  all the quantum numbers of the second nucleon. 

\begin{figure}[t]
\psfrag{k1}{$k_1$}
\psfrag{k2}{$k_2$}
\psfrag{k1-k}{$k_1-k$}
\psfrag{k2+k}{$k_2+k$}
\centerline{\epsfig{file=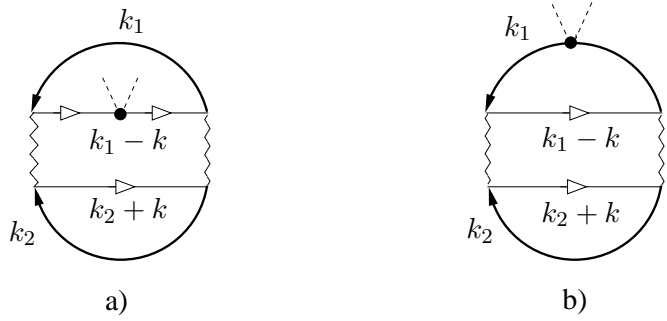,width=.5\textwidth,angle=0}}
\vspace{0.2cm}
\caption[pilf]{\protect \small
Contribution to the pion self-energy  with
a two-nucleon reducible loop. The
pion scatters inside/outside the loop for the diagram a)/b).
\label{fig:self1loop}}
\end{figure}

Proceeding similarly as before for $\Sigma_5$, including the integration by parts, we can write
\begin{align}
\Sigma_7&=\frac{q^0}{2f^2}\left(1-g_A^2\frac{\vq^2}{q_0^2}\right)\varepsilon_{ij3}\sum_{\sigma_1}\int\frac{d^4 k_1}{(2\pi)^4}\left(
G_0(k_1)_p\frac{\partial\Sigma_{p,NN}}{dk_1^0}-
G_0(k_1)_n\frac{\partial\Sigma_{n,NN}}{dk_1^0}
\right)e^{ik_1^0 \eta}~.
\label{sig.7.int}
\end{align}
Inserting in the previous equation the explicit expression for $\Sigma_{i_3,NN}$ of eq.\eqref{sel.n} one has
\begin{align}
\Sigma_7&=-\frac{iq^0}{2f^2} \left(1-g_A^2\frac{\vq^2}{q_0^2}\right) \varepsilon_{i j
3}\sum_{\sigma_1,\sigma_2}
 \int\frac{d^4k_1}{(2\pi)^4}\frac{d^4k_2}{(2\pi)^4}  e^{ik_1^0\eta}e^{ik_2^0\eta}\nn\\
 &\times \Biggl[G_0(k_1)_p G_0(k_2)_p \frac{\partial}{\partial
   k_1^0}T_{NN}(k_1 \sigma_1 p, k_2 \sigma_2 p|k_1 \sigma_1 p,
k_2\sigma_2 p)\nn\\
&- G_0(k_1)_n  G_0(k_2)_n \frac{\partial}{\partial
   k_1^0}T_{NN}(k_1 \sigma_1  n, k_2 \sigma_2 n|k_1 \sigma_1 n, k_2\sigma_2 n)\Biggr]~.
 \label{sigma.8}
\end{align}
This result is obtained from eq.\eqref{sig.7.int} by noting that
\begin{align}
\sum_{\sigma_1,\sigma_2}
 \int\frac{d^4k_1}{(2\pi)^4}\frac{d^4k_2}{(2\pi)^4}  e^{ik_1^0\eta}e^{ik_2^0\eta} &\Biggl[G_0(k_1)_p G_0(k_2)_n \frac{\partial}{\partial   k_1^0}T_{NN}(k_1 \sigma_1 p, k_2 \sigma_2 n|k_1 \sigma_1 p,
k_2\sigma_2 n)\nn\\
&- G_0(k_1)_n  G_0(k_2)_p \frac{\partial}{\partial
   k_1^0}T_{NN}(k_1 \sigma_1  n, k_2 \sigma_2 p|k_1 \sigma_1 n, k_2\sigma_2 p)\Biggr]=0~.
   \label{sig.7.can}
\end{align}
This follows  for two reasons. First, let us notice that because of  Fermi-Dirac statistics 
\begin{align}
T_{NN}(k_1 \sigma_1 p, k_2 \sigma_2 n|k_1 \sigma_1 p,k_2\sigma_2 n)=T_{NN}(k_2 \sigma_2 n, k_1 \sigma_1 p|k_2 \sigma_2 n,k_1\sigma_1 p)~.
\end{align}
Second, at LO the amplitude $T_{NN}$, as commented above, is given by the iteration of the wiggly line in fig.\ref{fig:nnself}.  The latter does neither depend on $k_1^0$ nor on $k_2^0$, see eqs.\eqref{feynman} and \eqref{1pi.gen}. Since  the two-nucleon reducible unitarity loop, the one in fig.\ref{fig:self1loop}b, depends on $k_1^0$ and $k_2^0$ through their sum, $k_1^0+k_2^0$, as can be seen straightforwardly \cite{longer}, it results that $T_{NN}$ at LO  only depends on them in this way. It follows then that $\partial T_{NN}/ \partial k_1^0=\partial T_{NN}/\partial k_2^0$. Taking these two facts into account, as  $k_i$ and $\sigma_i$ are dummy variables,   eq.\eqref{sig.7.can} results.

Let us now consider the diagrams c) and d) of fig.\ref{fig:all} whose
contribution is denoted by $\Sigma_8$. These diagrams consist of the
pion-nucleon scattering in a two-nucleon reducible loop which is corrected by
initial and final state interactions. The iterations are indicated by the
ellipsis on both sides of the diagrams.  
In order to see that these diagrams cancel with eq.(\ref{sigma.8})
let us take first the diagram of fig.\ref{fig:self1loop}a with a twice
iterated wiggly line vertex. It is given by

\begin{align}
\Sigma_8^{L}&=i\frac{q^0}{2f^2}\left(1-g_A^2\frac{\vq^2}{q_0^ 2}\right)\varepsilon_{ij3}\sum_{\alpha,\beta}\sum_{\sigma_1,\sigma_2}
\int\frac{d^4 k_1}{(2\pi)^4}\frac{d^4 k_2}{(2\pi)^4}
G_0(k_1)_\alpha G_0(k_2)_\beta e^{ik_1^0\eta}e^{ik_2^0\eta}\nn\\
&\times \Biggl[\frac{-i}{2}\int\frac{d^4q}{(2\pi)^4}\sum_{\alpha',\beta'}\sum_{\sigma'_1,\sigma'_2}V_{\alpha \beta;\alpha'\beta'}(q)\frac{\partial G_0(k_1-q)_{\alpha'}}{\partial k_1^0}\left.\tau_3\right|_{\alpha'\alpha'} V_{\alpha'\beta';\alpha\beta}(-q)G_0(k_2+q)_{\beta'}\Biggr]~,
\label{sig.8l.1}
\end{align}
where $V_{\alpha\beta;\gamma\delta}$ corresponds to the wiggly line   with the indices $\alpha$ and $\gamma$ belonging  to the out-/in-going first particle, in that order, and similarly $\beta$ and $\delta$ for the second one. To shorten the notation,  we have only indicated the isospin indices in $V$ in the previous equation, although $V$ depends also on spin. A symmetry factor $1/2$ is also included because $V_{\alpha\beta;\gamma\delta}$ contains both the direct and exchange terms,  as explicitly shown in eqs.\eqref{feynman} and \eqref{1pi.gen}. The
derivative with respect to $k_1^0$ arises in eq.(\ref{sig.8l.1}) because the
nucleon propagator to which the two pions are attached appears squared and we have made use of eq.\eqref{pro.sq}.  This is so because  for the $\pi^{\pm}$, $i$ and $j$ can be either 1 or 2, so that
the only surviving contribution is $k=3$. For the $\pi^0$, $i=j=3$ and then there
is no contribution.  Thus, because one has either 0 or $\tau^3$, which is a diagonal
matrix, the nucleon propagator before and after the two-pion vertex is the 
same. The $q$-loop integral in eq.\eqref{sig.8l.1} is typically divergent. Nevertheless, the parametric derivative with respect to $k_1^0$ can be extracted out of the integral as soon as it is regularized. Of course, the same regularization method as that used to calculate $T_{NN}$ should be employed.   Once the derivative is taken out of the integral, the quantity between the squared brackets in eq.\eqref{sig.8l.1} corresponds to the twice iterated wiggly line contribution to $T_{NN}(k_1\sigma_1\alpha,k_2\sigma_2\beta|k_1\sigma_1\alpha,k_2\sigma_2\beta)$.  In addition, the isovector nature of the modified WT vertex
of eq.(\ref{ver:wte}) implies that only the difference between the proton-proton
and neutron-neutron contributions arises. This can be worked out
straightforwardly from the isospin structure of the local four-nucleon  vertex 
and that of the one-pion exchange \cite{longer}. As a result we can write
\begin{align}
\Sigma_8^L&=\frac{iq^0}{2f^2}\left(1-g_A^2\frac{\vq^2}{q_0^2}\right)\varepsilon_{ij3}\sum_{\sigma_1,\sigma_2}\int\frac{d^4k_1}{(2\pi)^4}\frac{d^4k_2}{(2\pi)^4}
\Biggr[
G_0(k_1)_p G_0(k_2)_p\frac{\partial}{\partial k_1^0}T_{NN}^L(k_1 \sigma_1 p,k_2\sigma_2 p|k_1 \sigma_1 p,k_2\sigma_2 p)\nn\\
&-
G_0(k_1)_n G_0(k_2)_n\frac{\partial}{\partial k_1^0}T_{NN}^L(k_1 \sigma_1 n,k_2\sigma_2 n|k_1 \sigma_1 n,k_2\sigma_2 n)
\Biggl]e^{ik_1^0\eta}e^{ik_2^0\eta}~,
\label{sig.8l.2}
\end{align}
where the superscript $L$ on top of   $T_{NN}$ indicates that this scattering amplitude is calculated at the one-loop level. This result then cancels exactly with that of fig.\ref{fig:self1loop}b,  corresponding to the twice iterated wiggly line exchange
contribution to $T_{NN}$ in eq.(\ref{sigma.8}). This cancellation is explicit by reducing eq.\eqref{sigma.8} to the one-loop case for calculating $T_{NN}$.
Notice as well that the contribution to $T_{NN}$ given by the exchange
of only one wiggly line  
vanishes when inserted in
 eq.(\ref{sigma.8}) because it is independent of $k_1^0$, see eqs.\eqref{feynman} and \eqref{1pi.gen}.

\begin{figure}[t]
\psfrag{k1}{$k_1$}
\psfrag{k2}{$k_2$}
\psfrag{k1-k}{$k_1-k$}
\psfrag{k2+k}{$k_2+k$}
\centerline{\epsfig{file=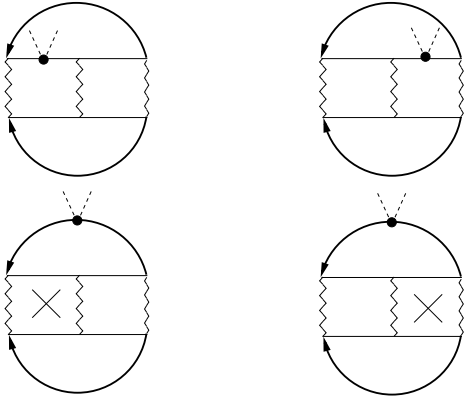,width=.35\textwidth,angle=0}}
\vspace{0.2cm}
\caption[pilf]{\protect \small
In this figure the cross indicates the action of the derivative with
respect to $k_1^0$ in eq.(\ref{sigma.8}). When the derivative is
performed over a baryon propagator the latter becomes squared, according
to eq.(\ref{pro.sq}). In this way, the first diagram on the second row
of the figure is the same as the one to the left of the first row but
with opposite sign and they cancel each other. The same applies to the
second diagrams on the first and second rows.
\label{fig:2loop}}
\end{figure}

This process of mutual cancellation  between $\Sigma_7$ and $\Sigma_8$ can be
generalized to any number of two-nucleon reducible loops in
figs.\ref{fig:all}a), b) and \ref{fig:all}c) and d), respectively. An $n+1$
iterated wiggly line exchange in these figures implies $n$
two-nucleon reducible loops. The two pions can be attached for
$\Sigma_8$ to any of them, while for $\Sigma_7$ the derivative with respect to
$k_1^0$ can also act on any of the loops. The iterative loops are the same for
both cases but  a relative minus sign results from the loop on which the two
pions are attached with respect to the one on which the derivative is acting, as just discussed.
This is exemplified in fig.\ref{fig:2loop} for the case with two two-nucleon
reducible loops. Hence,
\begin{align}
\Sigma_7+\Sigma_8=0~.
\label{can.89}
\end{align}
The basic simple reason for such cancellation is that 
while for $\Sigma_7$ the presence of a nucleon propagator squared gives rise to
$(-1)^2\partial/\partial k_1^0$, for $\Sigma_8$ it yields $-\partial/\partial
k_1^0$, cf. eqs.(\ref{sigma.8}) and (\ref{sig.8l.1}), respectively. The extra $(-1)$ for $\Sigma_7$ results because it involves an integration by parts, as discussed above. 
 
\section{Conclusions and outlook}
\def\theequation{\arabic{section}.\arabic{equation}}
\setcounter{equation}{0}
\label{sec:conc}

We have developed a promising scheme for an EFT in the nuclear medium that
combines both short-range and pion-mediated inter-nucleon interactions. It is
based on the development of a new chiral power counting which is bounded from
below and at a given order it requires to calculate a finite number of
contributions.  The latter could eventually involve  infinite strings of 
two--nucleon reducible diagrams with the leading ${\cal O}(p^0)$ two-nucleon CHPT amplitudes. As a result, our power counting accounts for non-perturbative effects to be
resummed which,  e.g., give rise to the generation of the deuteron in vacuum
nucleon-nucleon scattering. The power counting from the onset takes into account
the presence of enhanced nucleon propagators and it can also be applied to
multi-nucleon forces.

We have then calculated the leading corrections to the lowest
order result for the pion self-energy in asymmetric nuclear matter, with all the  contributions up-to-and-including ${\cal O}(p^5)$  evaluated. As a novelty, it
is shown that the leading corrections to the linear density approximation vanish. In particular, 
it is derived that the leading corrections from nucleon-nucleon scattering mutually cancel. This suppression is interesting since it allows to understand from first principles the phenomenological success of fitting data on pionic atoms with only meson-baryon interactions \cite{weiprl,galrep}.    
An ${\cal O}(p^6)$ calculation of the pion self-energy is a very
interesting task as it provides the first corrections to the linear density approximation, e.g.  the well-known Ericson-Ericson-Pauli rescattering effect \cite{ericeric}. 
More calculations for other physical processes and higher orders are 
clearly needed to assess
the realm of applicability of the present approach.

\section*{Acknowledgements}
We would like to thank  Andreas~Wirzba for discussions and encouragement. J.A.O. also thanks Eulogio~Oset for informative discussions.
This work is partially funded by the grant MEC  FPA2007-6277 and by 
the BMBF grant 06BN411,  EU-Research Infrastructure
Integrating Activity
 ``Study of Strongly Interacting Matter" (HadronPhysics2, grant n. 227431)
under the Seventh Framework Program of EU 
and HGF grant VH-VI-231 (Virtual Institute ``Spin and strong QCD'').



\begin{thebibliography}{99}
\bibitem{wein}{ S.~Weinberg,} 
Physica A {\bf96} (1979) 327.
\vs
\bibitem{wein1}  S.~Weinberg,
  Phys.\ Lett.\  B {\bf 251} (1990)  288.
  \vs
\bibitem{wein2}  S.~Weinberg,
  Nucl.\ Phys.\   B {\bf 363} (1991) 3.
\vs
\bibitem{ordo} C.~Ordonez, L.~Ray and U.~van Kolck,
  Phys.\ Rev.\  C {\bf 53}  (1996) 2086.
\vs
\bibitem{kolck}U.~van Kolck,
  Prog.\ Part.\ Nucl.\ Phys.\  {\bf 43}  (1999) 337.
\vs
\bibitem{entem} D.~R.~Entem and R.~Machleidt,
  Phys.\ Rev.\  C {\bf 68}  (2003) 041001.
\vs
\bibitem{epe}
E.~Epelbaum, W.~Gl\"ockle and U.-G.~Mei\ss ner,
 Nucl.\ Phys.\  A {\bf 671}  (2000) 295;  Nucl.\ Phys.\  A {\bf 747}  (2005) 362.
\vs
\bibitem{epeprl}  E.~Epelbaum, H.~Kamada, A.~Nogga, H.~Witala, W.~Gl\"ockle and U.-G.~Mei\ss ner,
  Phys.\ Rev.\ Lett.\  {\bf 86}  (2001) 4787.
\vs
\bibitem{eperp}E.~Epelbaum,
  Prog.\ Part.\ Nucl.\ Phys.\  {\bf 57}  (2006) 654.
\vs
\bibitem{Epelbaum:2008ga}
  E.~Epelbaum, H.~W.~Hammer and U.-G.~Mei{\ss}ner, {\it Rev. Mod. Phys.}, to appear,
  arXiv:0811.1338 [nucl-th].
\vs
\bibitem{schaefer} R.~J.~Furnstahl, G.~Rupak and T.~Sch\"afer,
Ann.\ Rev.\ Part.\ Nucl.\ Sci.\ {\bf 58} (2008) 1.

\vs
\bibitem{nogga} S.~K.~Bogner, R.~J.~Furnstahl, S.~Ramanan and A.~Schwenk,
  Nucl.\ Phys.\  A {\bf 773}  (2006) 203;   S.~K.~Bogner, A.~Schwenk, R.~J.~Furnstahl and A.~Nogga,
  Nucl.\ Phys.\  A {\bf 763}  (2005) 59.
\vs
\bibitem{kaiser}  N.~Kaiser, M.~M\"uhlbauer and W.~Weise,
  Eur.\ Phys.\ J.\  A {\bf 31}  (2007) 53.
\vs
\bibitem{epenm} P.~Saviankou, S.~Krewald, E.~Epelbaum and U.-G.~Mei\ss ner,
  arXiv:0802.3782 [nucl-th].
\vs
\bibitem{Navratil:2007we}
  P.~Navratil, V.~G.~Gueorguiev, J.~P.~Vary, W.~E.~Ormand and A.~Nogga,
  Phys.\ Rev.\ Lett.\  {\bf 99} (2007) 042501.
\vs
\bibitem{prcoller} J.~A.~Oller,
  Phys.\ Rev.\  C {\bf 65}  (2002) 025204.
\vs
\bibitem{annp}  U.-G.~Mei\ss ner, J.~A.~Oller and A.~Wirzba,
  Annals Phys.\  {\bf 297}  (2002) 27.
\vs
\bibitem{Girlanda:2003cq}
  L.~Girlanda, A.~Rusetsky and W.~Weise,
  Annals Phys.\  {\bf 312} (2004) 92.
\vs
\bibitem{kai1}  N.~Kaiser, S.~Fritsch and W.~Weise,
  Nucl.\ Phys.\ A  {\bf 697}  (2002) 255.
\vs
\bibitem{kai2} N.~Kaiser, S.~Fritsch and W.~Weise,
  Nucl.\ Phys.\  A {\bf 724}  (2003) 47.
\vs
\bibitem{kai3} S.~Fritsch, N.~Kaiser and W.~Weise,
  Nucl.\ Phys.\  A {\bf 750}  (2005) 259.
  \vs
\bibitem{hardrock}  R.~Rockmore,
  Phys.\ Rev.\  C {\bf 40}  (1989) 13.
\vs
\bibitem{osetdo} M.~D\"oring and E.~Oset,
  Phys.\ Rev.\   C {\bf77}   (2008) 024602.
\vs
\bibitem{osetnie} E.~Oset, C.~Garcia-Recio and J.~Nieves,
  Nucl.\ Phys.\  A {\bf 584}  (1995) 653.
\vs
\bibitem{korean} 
T.~S.~Park, H.~Jung and D.~P.~Min,
  J.\ Korean Phys.\ Soc.\  {\bf 41}  (2002) 195.
\vs
\bibitem{goldstone}  J.~Goldstone, A.~Salam and S.~Weinberg,
  Phys.\ Rev.\  {\bf 127} (1962) 965.
\vs
\bibitem{bern} J.~Gasser and H.~Leutwyler,
  Annals Phys.\  {\bf 158} (1984) 142;  Nucl.\ Phys.\  B {\bf 250} (1985) 465.
\vs
\bibitem{jan} N.~H.~Fuchs, H.~Sazdjian and J.~Stern,
  Phys.\ Lett.\  B {\bf 269} (1991) 183; 
  Phys.\ Rev.\  D {\bf 47} (1993) 3814.
\vs
\bibitem{prlbern} G.~Colangelo, J.~Gasser and H.~Leutwyler,
  Phys.\ Rev.\ Lett.\  {\bf 86} (2001) 5008.
\vs
\bibitem{andreas} V.~Thorsson and A.~Wirzba, Nucl. Phys. A {\bf 589} (1995) 633;  ''Hirschegg '95: Dynamical Properties of Hadrons in Nuclear Matter" (H. Feldmeier and W.N\"orenberg,   Eds.), pp. 31-43,  GSI-print, Darmstadt, 1995; M.~Kirchbach and A.~Wirzba, Nucl. Phys. A {\bf 604} (1996) 395; {\it ibid} A {\bf 616} (1997) 648. 

\vs
\bibitem{condenk}
  N.~Kaiser and W.~Weise,
  Phys.\ Lett.\  B {\bf 671} (2009) 25.
\vs
\bibitem{ericeric}M. Ericson and T.~E.~O. Ericson, Annals Phys.\  {\bf 36}  (1966) 323.
\vs
\bibitem{galrep}  E.~Friedman and A.~Gal,
  Phys.\ Rept.\  {\bf 452}  (2007) 89, and references therein.
\vs
\bibitem{chanfray}  G.~Chanfray, M.~Ericson and M.~Oertel,
  Phys.\ Lett.\  B {\bf 563}  (2003) 61.
\vs
\bibitem{kwpk} N.~Kaiser and W.~Weise,
  Phys.\ Lett.\ B  {\bf 512}  (2001) 283.
\vs
\bibitem{Girlanda:2004qa}
  L.~Girlanda, A.~Rusetsky and W.~Weise,
  Nucl.\ Phys.\  A {\bf 755} (2005) 653.
\vs
\bibitem{weiprl} E.~E.~Kolomeitsev, N.~Kaiser and W.~Weise,
  Phys.\ Rev.\ Lett.\  {\bf 90}  (2003) 092501.
\vs
\bibitem{original} J.~D.~Walecka,
  Annals Phys.\  {\bf 83} (1974) 491.
\vs
\bibitem{rev1}B.~D.~Serot and J.~D.~Walecka,
  Adv.\ Nucl.\ Phys.\  {\bf 16} (1986) 1.
\vs
\bibitem{rev2} B.~D.~Serot,
  Rept.\ Prog.\ Phys.\  {\bf 55}, 1855 (1992).
\vs
\bibitem{short1} B.~D.~Serot and J.~D.~Walecka,
  arXiv:nucl-th/0010031.
\vs
\bibitem{short2}R.~J.~Furnstahl and B.~D.~Serot,
  Comments Nucl.\ Part.\ Phys.\  {\bf 2} (2000) A23.
\vs
\bibitem{bdserots} J.~A.~McNeil, J.~R.~Shepard and S.~J.~Wallace,
  Phys.\ Rev.\ Lett.\  {\bf 50} (1983) 1439;  B.~C.~Clark, S.~Hama, R.~L.~Mercer, L.~Ray and B.~d.~Serot,
  Phys.\ Rev.\ Lett.\  {\bf 50} (1983) 1644.
\vs
\bibitem{sainio}J.~Gasser, M.~E.~Sainio and A.~Svarc,
  Nucl.\ Phys.\ B  {\bf 307} (1988)779.
\vs
\bibitem{Borasoy:2006qn}
  B.~Borasoy, E.~Epelbaum, H.~Krebs, D.~Lee and U.-G.~Mei{\ss}ner,
  Eur.\ Phys.\ J.\  A {\bf 31} (2007) 105.
\vs
\bibitem{fetter}A.~L.~Fetter and J.~D.~Walecka, 
``Quantum Theory of Many-Particle Systems''.  Dover Publications, Inc., Mineola, New York, 2003.
\vs
\bibitem{nnk} N.~Kaiser, R.~Brockmann and W.~Weise,
  Nucl.\ Phys.\  A {\bf 625} (1997) 758
\vs
\bibitem{ulfrev} V.~Bernard, N.~Kaiser and U.-G.~Mei\ss ner,
  Int.\ J.\ Mod.\ Phys.\  E \ {\bf 4}  (1995) 193. 
\vs
\bibitem{longer}
  A.~Lacour, J.~A.~Oller and U.-G.~Mei{\ss}ner,
  ``Non-perturbative methods for a chiral effective field theory of finite density nuclear systems,''
  arXiv:0906.2349 [nucl-th].
\end{thebibliography}
\end{document}